\documentclass{article}

\usepackage{amssymb,amsfonts,amsmath}
\usepackage{cite,enumerate,float}
\usepackage{color}
\usepackage{tikz}
\usetikzlibrary{arrows,snakes,backgrounds}
\usepackage{url}

\usepackage[utf8]{inputenc}
\usepackage[russian]{babel}

\def\be{\begin{eqnarray}}
\def\ee{\end{eqnarray}}
\def\nn{\nonumber}

\def\p{\partial}

\def\Tr{{\rm Tr}\,}

\newcommand{\ttop}[1]{
  q^{\hat{D}_#1}
}

\definecolor{red}{rgb}{1,0,0}
\definecolor{orange}{rgb}{1,0.5,0}
\definecolor{violet}{rgb}{0.7,0,1}

\def\sgn{\hbox{sgn}}
\def\e{{\bf\mathfrak{e}}}

\hoffset= - 1.0in         

\begin{document}

\title{\vspace{1.5cm}\bf
Non-commutative creation operators for symmetric polynomials
}

\author{
A. Mironov$^{b,c,d,}$\footnote{mironov@lpi.ru,mironov@itep.ru},
A. Morozov$^{a,c,d,}$\footnote{morozov@itep.ru}
}

\date{ }

\maketitle

\vspace{-6.5cm}

\begin{center}
FIAN/TD-14/25  \hfill {\bf to the memory of}\\
ITEP/TH-31/25   \hfill {\bf Masatoshi Noumi}\\
IITP/TH-28/25  \hfill \phantom{.}\\
MIPT/TH-23/25 \hfill \phantom{.}
\end{center}

\vspace{4.5cm}

\begin{center}
$^a$ {\small {\it MIPT, Dolgoprudny, 141701, Russia}}\\
$^b$ {\small {\it Lebedev Physics Institute, Moscow 119991, Russia}}\\
$^c$ {\small {\it NRC ``Kurchatov Institute", 123182, Moscow, Russia}}\\
$^d$ {\small {\it Institute for Information Transmission Problems, Moscow 127994, Russia}}
\end{center}

\vspace{.1cm}

\begin{abstract}
We reconsider in modern terms the old discovery by A. Kirillov and M. Noumi,
who devised peculiar operators adding columns to Young diagrams enumerating the Schur, Jack and Macdonald polynomials.
In this sense, these are a kind of ``creation'' operators, representing Pieri rules in a maximally simple form,
when boxes are added to Young diagrams in a regular way and not to arbitrary ``empty places'' around the diagram.
Instead the operators do not commute, and one should add columns of different lengths one after another.
We consider this construction in different contexts. In particular,
we build up the creation operators $\hat B_m$ in the matrix and Fock representations of the $W_{1+\infty}$ algebra, and in the Fock representation of the affine Yangian algebra $Y(\widehat{gl}_1)$.
\end{abstract}

\bigskip

\section{Introduction}

Wave functions in the systems of particles with coordinates $\vec x_i$ are often made from the polynomials
of these variables or their simple functions.
In general, these wave functions depend on permutations $w$ of the coordinates and form a variety $\Big\{P_w[\vec x]\Big\}$.
Different physical systems are associated with different bases in this variety,
and their symmetries, with action of different groups on it.
Theory of such polynomials is vast, but not very well structured, because of the lack of clear classification principles.

In recent decades a partial order was introduced by restriction to integrable systems,
where the special role is played by one-dimensional (perhaps, complex) coordinates
and symmetric polynomials, which are labeled by Young diagrams (partitions) $\lambda$ instead of arbitrary permutations $w$.
Moreover, there are special bases, provided by
symmetric polynomials from the Schur-Jack-Macdonald family \cite{Mac}.
After discovery of the role of integrability \cite{UFN3} and superintegrability  \cite{SIrev,SIU,MMNek}
in non-perturbative physics,
attention to these polynomials and their application to variety of physical problems increased drastically,
and so did the need for a systematic theory, which is still lacking.

Originally, the Schur polynomials were related to characters of highest weight representations of $SL_N$ groups \cite{schurchar},
then they could be considered as building blocks of the eigenfunctions of commuting Hamiltonians
in the trigonometric Calogero-Sutherland systems \cite{Cal,Suth,OP} in the free fermion point,
and acquired extension to Jack \cite{Jack,St,Turb,GP} and Macdonald \cite{Mac} polynomials , which are beyond the representation theory of Lie algebras.
Nowadays they are related to a wide system of commutative subalgebras \cite{Ch1,Ch2,Ch3,MMMP1,MMMP2,MMP} of the $W_{1+\infty}$ \cite{Pope,FKN2,BK,BKK,KR1,FKRN,Awata,KR2}, affine Yangian \cite{SV,AS,Tsim,Prochazka} and quantum toroidal/Ding-Iohara-Miki (DIM) \cite{DI,Miki}\footnote{Or the elliptic Hall algebra \cite{K,BS,S}, which is basically the same \cite{S,Feigin}.} algebras.

Cut-and-join operators of \cite{GJ,MMN} form the simplest of such subalgebras (``vertical ray''), but there are many more,
labeled by all kinds of ``rays'' and ``cones'' \cite{MMMP1}.
These algebras have many different representations, accordingly symmetric polynomials can be represented differently.
In particular, in the case of Lie algebra $W_{1+\infty}$, there is a representation in terms of $N\times N$ matrices \cite{MMCal,MMMP1}, where the coordinates $x_i$ are considered as matrix eigenvalues. At the same time, the $N$-body representation in terms of $x_i$'s can be constructed for all these three algebras, and so does the Fock representation. In the latter case, the symmetric polynomials of $x_i$ are
realized as polynomials of the power sums $p_k=\sum_{i=1}^N x_i^k$.
All three algebras also admit the large MacMahon representation, when the Young diagrams become plane partitions \cite{Tsim,Prochazka,Jimbo}.
Polynomials associated with different diagrams are connected by Pieri rules, which allow one to add and subtract boxes,
but the rules are somewhat sophisticated, because the boxes can not be added at arbitrary places.
Moreover, the action of standard generators add boxes at rather arbitrary places.
The Pieri rules can be considered as an alternative (dual) description of algebras, defining there action on
various systems of Young diagrams and their generalizations.
For recent reviews and further elaboration on the structure and use of the Pieri rules within the described framework, see \cite{MMZ,GMTsPieri,AT}.

This paper is devoted to a description
of a spectacular old observation about modification of the Pieri rules, which puts them in a nice, though unusual order.
That is, in the seminal \cite{KN}, A. Kirillov and M. Noumi introduced an amusing kind of creation and annihilations operators for the
Schur, Jack and Macdonald symmetric polynomials (similar operators were constructed earlier in \cite{Vinet} for the first two of these polynomial systems).
The creation operators $\hat B_m$ adds the column $[1^m]$ to the Young diagram labeling the symmetric polynomial if this Young diagram does not contain columns of the length more than $m$, and the polynomial associated with the Young diagram
$R=\{r_1\geq r_2\geq \ldots r_{n-1}\geq r_n\}$ is realized by these operators as
\be
P_\lambda = \hat B_n^{r_n} \hat B_{n-1}^{r_{n-1}-r_n} \ldots \hat B_1^{r_1-r_2}\cdot\underbrace{P_\emptyset}_{=const}
\ee
i.e.
\be
P_{[r]} &=& \hat B_1^r\cdot P_\emptyset,
\nn \\
P_{[r_1,r_2]} &=& \hat B_2^{r_2} \hat B_1^{r_1-r_2}\cdot P_\emptyset,
\nn \\
P_{[r_1,r_2,r_3]} &=& \hat B_3^{r_3}\hat B_2^{r_2-r_3}\hat B_1^{r_1-r_2}\cdot P_\emptyset,
\nn \\
\ldots
\ee

In \cite{KN} these operators were built in terms of the Dunkl and Cherednik operators acting on symmetric functions of $N$ variables $x_i$, i.e. in the $N$-body representation of the corresponding affine Yangian and quantum toroidal algebras \cite{MMP} (equivalently, one can interpret them in terms of DAHA \cite{Ch}).
In this paper we devise these operators in the
matrix and Fock representations, i.e. in terms of matrices $\Lambda$ (like the ones used in \cite{MMN})
and in terms of power sums $p_k=\sum_{i=1}^N x_i^k$.

An important advantage of the Fock representation is that the polynomials from the Schur-Jack-Macdonald family when considered as polynomials of power sums have a simple behaviour \cite{Mac} under the transposition of the Young diagram $R\to R^\vee$:
\be
S_R(p_k)=S_{R^\vee}\Big((-1)^{k+1}p_k\Big)\nn\\
J_R(p_k;\beta)\sim J_{R^\vee}\Big((-1)^{k+1}\beta p_k;\beta^{-1}\Big)\nn\\
M_R(p_k;q,t)\sim
M_{R^\vee}\left((-1)^{k+1}{1-t^k\over 1-q^k}p_k;t,q\right)
\ee
where $S_R(p_k)$, $J_R(p_k)$, $M_R(p_k)$ are the Schur, Jack and Macdonald polynomials respectively.
Therefore, one can immediately make from the operators $\hat B_m$ the operators adding lines instead of columns by a simple rescaling of $p_k$'s.

The paper is organized as follows. In section 2, we briefly review some results of \cite{KN} who obtained the creation operators in the $N$-body representation, and explain simple steps that give rise to these operators in the Fock representation, we also discuss make a few comments about constructing there operators in section 3. Since section 4, we start a systematic description of the creation operators $\hat B_m$, and build them up in the matrix representation of $W_{1+\infty}$ algebra, when the operators generate the Schur polynomials. We explain how one can convert these formulas in the matrix representation into those in the Fock representation in section 5, while section 6 contains the complete description of the operators in the Fock representation. Section 7 contains an extension of these results to the Fock representation of the affine Yangian algebra $Y(\widehat{gl}_1)$, , when the operators generate the Jack polynomials. Some concluding comments are contained in section 7.

\paragraph{Notation.} In the paper, we deal with symmetric polynomials of $N$ variables $x_i$ as with graded polynomials of power sums $p_k=\sum_ix_i^k$. The polynomials that we discuss are the Schur polynomials $S_R(p_k)$, and their deformations to the Jack polynomials $J_R(p_k)$ and Macdonald polynomials $M_R(p_k)$, all of them being enumerated by the Young diagrams (partitions) $R$ \cite{Mac}. A special role in the construction is played by the elementary symmetric polynomials
\be
\e_n[x_i]=\sum_{1\le i_1<i_2<\ldots<i_n}x_{i_1}x_{i_2}\ldots x_{i_n}\le N
\ee
These polynomials coincide with all above (Schur, Jack and Macdonald) symmetric polynomials at $R=[1^n]$ at once, since the Schur polynomial $S_{[1^n]}$ is not deformed.

In terms of power sums, the elementary symmetric polynomials are given by the generating function
\be
\exp\left(\sum_k(-1)^{k+1}p_k{z^k\over k}\right)=\sum_n\e_n(p_k) z^n
\ee
They can be also written in this case as
\be
\e_n(p_k)={1\over n!}\sum_{\sigma\in{\cal S}_n}\sgn(\sigma)\prod_jp_{l(\sigma_j)}
\ee
where the sum runs over all permutations $\sigma$ of the symmetric group ${\cal S}_n$, and $l(\sigma_j)$ is the length of the $j$-th cycle if the permutation $\sigma$ is realized by a set of cycles $\sigma_j$ in disjoint cycle notation.

The first examples of these polynomials, which we use in examples in this paper are
\be
\e_1(p_k)&=&p_1\nn\\
\e_2(p_k)&=&{p_1^2\over 2}-{p_2\over 2}\nn\\
\e_3(p_k)&=&{p_1^3\over 6}-{p_1p_2\over 2}+{p_3\over 3}\nn\\
\e_4(p_k)&=&{p_1^4\over 24}-{p_2p_1^2\over 4}+{p_3p_1\over 3}+{p_1^2\over 8}-{p_4\over 4}\nn\\
\e_4(p_k)&=&{p_1^5\over 120}-{p_2p_1^3\over 12} +{p_3p_1^2\over 6}+{p_2^2p_1\over 8}-{p_4p_1\over 4}-{p_3p_2\over 6}+{p_5\over 5}
\ee

\section{Original $N$-body representation}

We start with the operators $\hat B_m$ acting to the Jack polynomials as symmetric functions of $N$ variables $x_i$. These polynomials are associated with the $N$-body representation of the affine Yangian algebra $Y(\widehat{gl}_1)$.
According to \cite{KN},
\be
\hat B_m = \sum_{1\le k_1<\ldots<k_m\le N}  x_{k_1}\ldots x_{k_m}(\hat D_{k_1}+\beta)(\hat D_{k_2}+2\beta)\ldots (\hat D_{k_m}+m\beta)
\label{KNops}
\ee
with the Dunkl operator \cite{Dunkl}
\be
\hat D_i = \underbrace{x_i\frac{\p}{\p x_i}}_{\hat\Lambda_i} + \beta\sum_{j\neq i} \frac{x_i}{ x_i-x_j}(1-\hat\sigma_{ij} )
\ee
where $\hat\sigma_{ij}$ permutes $x_i$ and $x_j$.

When acting on symmetric functions (which is denoted by $\approx$),
\be
\hat B_1 = \sum_i x_i(\hat D_i+\beta) \approx \sum_i x_i(\hat\Lambda_i + \beta)
= \beta p_1 + \sum_a ap_{a+1}\frac{\p}{\p p_a}
\ee
since $(1-\hat\sigma_{ij})$ annihilates them. Here $p_k=\sum_ix_i^k$.

However, for the next operator, this is no longer so simple: the operator $(1-\hat\sigma_{ij})$ acts non-trivially even on symmetric functions. It is only the special adjustment of the coefficients that the net result remains symmetric:
\be
\hat B_2 =\sum_{i<j} x_i x_j(\hat D_i +\beta)(\hat D_j+2\beta)
\approx \sum_{i<j} x_ix_j \left(\hat \Lambda_i
+\beta\sum_{k\neq i} \frac{x_i(1-\hat\sigma_{ik} )}{x_i-x_k} + \beta\right)(\hat\Lambda_j + 2\beta)
\approx \nn \\
\approx \sum_{i<j} \left\{x_ix_j\hat\Lambda_i\hat\Lambda _j + 2\beta^2 x_ix_j
+ \beta x_ix_j\left( \hat\Lambda_j + 2\hat\Lambda_i + \frac{x_i}{x_i-x_j}(\hat\Lambda_j-\hat\Lambda_i)\right)\right\}
\label{B2inx}
\ee
Note that only $k=j$ contributes to the sum, when the action is on symmetric functions.
At the first glance, the $\beta$-linear term at the r.h.s. does not look symmetric.
However, it is such, being equal to
\be
\frac{\beta x_ix_j}{x_i-x_j}\left( (x_i-x_j)(  \hat\Lambda_j+2\hat\Lambda_i ) + x_i(\hat\Lambda_j-\hat\Lambda_i)\right)
=  \frac{\beta x_ix_j}{x_i-x_j}\left( (x_i-2x_j)\hat\Lambda_i - (x_j-2x_i)\hat\Lambda_j  \right)
\label{B2betalinear}
\ee
The $\beta^2$ term in (\ref{B2inx}) is just $\beta^2(p_1^2-p_2)$, while the other terms reproduce formula (\ref{Bvsp}) below after conversion to the $p$-variables (power sums).
Indeed,
\be
\sum_{i<j} x_ix_j \hat\Lambda_i\hat\lambda_j  =
 \sum_{a,b} \sum_{i<j} x_i^{a+1}x_j^{a+1} ab\frac{\p^2}{\p p_a\p p_b}
 =  \frac{1}{2}\sum_{a,b}ab\left(p_{a+1}p_{b+1} - p_{a+b+2}\right)\frac{\p^2}{\p p_a\p p_b}
\ee
while (\ref{B2betalinear}) is
\be
 \beta \sum_c  \left\{\sum_{i<j}\frac{x_ix_j}{x_i-x_j}\Big(  (x_i-2x_j)x_i^c-(x_j-2x_i)x_j^c\Big)\right\}   c\frac{\p}{\p p_c}
= \nn \\
 = \frac{\beta}{2}   \sum_c  \left\{\sum_{i,j}\left(
x_ix_j\frac{x_i^{c+1} - x_j^{c+1}}{x_i-x_j} -2 x_i^2x_j^2 \frac{x_i^{c-1}-x_j^{c-1}}{x_i-x_j}\right)
-\sum_i\Big((c+1)-2(c-1)\Big)x_i^{c+2}\right\}
c\frac{\p}{\p p_c}
= \nn\\
  = \frac{\beta}{2}\left\{\sum_{a,b\geq 1}\left( (a+b-2)p_ap_b\frac{\p}{\p p_{a+b-2}} -2(a+b) p_{a+1}p_{b+1}\frac{\p}{\p p_{a+b}} \right)
+ \sum_{a\geq 1} a(a-3)p_{a+2}\frac{\p}{\p p_a}\right\}
= \nn \\
= \frac{\beta}{2}\left\{-\sum_{a,b} (a+b-2)p_ap_b\frac{\p}{\p p_{a+b-2}}
+ 4p_1 \sum_a ap_{a+1}\frac{\p}{\p p_a}
+ \sum_a a(a-3)p_{a+2}\frac{\p}{\p p_a}\right\}
\ee

The same trick works for the higher $\hat B_m$, though in a non-trivial way, but (\ref{KNops}) preserves
symmetric functions and can be rewritten in terms of $p$-variables.

In the case of Macdonald deformation, one should substitute the Dunkl operators $\hat D_i$ by the Cherednik ones \cite{Ch}. Another way to explicitly realize $\hat B_m$ when acting on symmetric functions in the Macdonald case is to use formulas (2), (6) from \cite{KN}, we reproduce them in Appendix A.

\section{Creation operators in the Fock representation}

\paragraph{First creation operators.}
In $p_k$ variables, the operators $\hat B_m$ look as follows:
\be
\hat B_1 &=& \beta p_1 + \sum_a ap_{a+1}\frac{\p}{\p p_a},
\nn \\
\hat B_2 &=& \beta^2 (p_1^2-p_2) + 2\beta p_1\sum_a ap_{a+1}\frac{\p}{\p p_a} + \frac{\beta}{2}\sum_a a(a-3)p_{a+2}\frac{\p}{\p p_a}
+ \beta\sum_{a,b} (a+b-2)p_ap_b \frac{\p}{\p p_{a+b-2}} + \nn\\
&+&\frac{1}{2}\sum_{a,b} ab\left(p_{a+1}p_{b+1}-p_{a+b+2}\right)\frac{\p^2}{\p p_a\p p_b},
\nn \\
\ldots
\label{Bvsp}
\ee
Note that these ``creation operators''  $\hat B_m$ with different $m$ do not commute.

The Macdonald deformation is
\be
\hat{\cal B}_1 &=& \frac{t-1}{q-1}p_1 + \sum_a \frac{(t^{a+1}-1)(q^a-1)}{(t^a-1)(q^{a+1}-1)}\cdot ap_{a+1}\frac{\p}{\p p_a}\sim\nn\\
&\sim&
\oint {dz\over z^2}\exp\left(\sum_k(1-t^{-k}){z^kp_k\over k}\right)\exp\left(\sum_k(1-q^k)z^k{\p\over\p p_k}\right)+{1-t\over t^2}p_1
\nn \\
\ldots
\label{MacB1}
\ee
$\hat{\cal B}_1$ is nothing but the element $e_{[1,1]}$ of the elliptic Hall algebra in the Fock representation.
Note that $a$ in $a\frac{\p}{\p p_a}$ is not deformed, which is typical for this deformation.
The Jack polynomial case is reproduced in the limit of $q,t\rightarrow 1$ with $t=q^\beta$.

\paragraph{Using cut-and-join operators.} The creation operators (\ref{Bvsp}) can be also realized in a slightly different way. In order to see this, we use the lowest cut-and-join operators \cite{GJ,MMN}:
\be
\hat\psi_2 &=& \sum_a ap_a \frac{\p}{\p p_a}, \nn \\
\hat\psi_3 &=& \frac{1}{2}\sum_{a,b}\left(abp_{a+b} \frac{\p^2}{\p p_a\p p_b} + \beta(a+b)p_ap_b\frac{\p}{\p p_{a+b}}\right)
-\frac{\beta-1}{2}\sum_a a(a-1)p_a\frac{\p}{\p p_a}
\label{psi23}
\ee
They are, in fact, the first generators $\psi_i$ of the affine Yangian algebra in the basis of \cite[sec.2.1]{Prochazka}, hence the notation.

The Jack polynomials are the eigenfunctions of these operators:
\be
\hat\psi_2 P_\lambda = |\lambda| P_\lambda, \nn \\
\hat\psi_3 P_\lambda = \epsilon_\lambda P_\lambda
\ee
The first creation operator can be then rewritten as
\be
\hat B_1 =\beta p_1 + [\hat \psi_3,p_1]
\label{B1viapsi}
\ee
and the other operators as
\be\label{Bmp}
\hat B_m = \frac{m(m+1)}{2}\beta^2 \e_m (p_k) +  \beta [\hat \psi_3, \e_m(p_k)] + \hat b_m
\ee
where the ``correction'' operator $\hat b_m$ annihilates all polynomials in the symmetric representations:
\be
\hat b_m P_{[r]}=0,
\label{annih}
\ee
i.e.
\be
(r+m\beta)P_{[r+1,1^{m-1}]} =   \left(\frac{m(m+1)}{2}\beta \e_m(p_k)  + [\hat \psi_3, \e_m(p_k)]\right)P_{[r]}
= \nn \\
= \left( \frac{m(m+1)}{2}\beta \e_m(p_k)  +  [\hat \psi_3,\e_m(p_k)]\right)
 \Big( \beta p_1  + [\hat \psi_3, p_1]\Big)P_0
\ee
However, generation of  $P_{[r+k,k^{m-1}]}$ with $k>1$ requires the knowledge of $\hat b_m$.

At $m=1$, there is no operator that annihilates all $P_{[r]}$, thus $\hat b_1 =0$.

At $m=2$, there are two operators:
\be
\hat a_2^{(1)}:=\frac{1}{2}\sum_{a,b=1} abp_{a+b+2}\frac{\p^2}{\p p_a\p p_b} - \frac{\beta}{2}\sum_{a=1} a(a-1)p_{a+2}\frac{\p}{\p p_a}
\ee
and
\be
\hat a_2^{(2)}:=\frac{1}{2}\sum_{a,b=1} abp_{a+1}p_{b+1}\frac{\p^2}{\p p_a\p p_b}
\underbrace{-\frac{\beta}{2}\sum_{a,b=1} (a+b-2)p_ap_b\frac{\p}{\p p_{a+b-2}} + \beta p_1 \sum_a ap_{a+1}\frac{\p}{\p p_a}}_{
-\frac{\beta}{2}\sum_{a,b=1} (a+b)p_{a+1}p_{b+1}\frac{\p}{\p p_{a+b}}}
\ee
The ``correction'' operator is their particular linear combination:
\be
\hat b_2 = \hat a_2^{(2)}-\hat a_2^{(1)}
\label{b2}
\ee
and
\be
(r_1+2\beta)(r_2+\beta) J_{[r_1+1,r_2]} = \Big(3\beta^2\e_2(p_k)+\beta[\hat \psi_3,\e_2(p_k)]
-\hat a_2^{(1)}+\hat a_2^{(2)} \Big)J_{[r_1,r_2]}
\ee

\section{Matrix representation\label{mr}}

So far, we worked with the Jack polynomials associated with the affine Yangian algebra $Y(\widehat{gl}_1)$. The Schur polynomials associated with the $W_{1+\infty}$ algebra are obtained at the particular value of $\beta=1$.

In order to deal with the Fock representation, it turns out to be technically convenient to start with the matrix representation first. This option is available for the $W_{1+\infty}$ algebra, which has the matrix representation \cite{MMMP1}. Since the case of $W_{1+\infty}$ algebra is associated with the Schur polynomials, which are graded polynomials of degrees of an $N\times N$ matrix, $\Tr\Lambda^k$ when considering the matrix representation, we deal with the Schur polynomial case in the next three sections, and define the matrix derivative operator\footnote{Hereafter, by the matrix derivative, we imply the derivative w.r.t. matrix elements of the transposed matrix: $\left(\frac{\partial}{\partial \Lambda}\right)_{ij}=\frac{\partial}{\partial \Lambda_{ji}}$.} $D_\Lambda=\Lambda {\p\over\p\Lambda}$,
obtaining
\be
B_1&=&\e_1\Big(\Tr\Lambda^k\Big)+\Tr(\Lambda D_\Lambda)
\nn\\
B_2&=&2\e_2\Big(\Tr\Lambda^k\Big)-\Tr(\Lambda^2 D_\Lambda)+\e_1\Big(\Tr\Lambda^k\Big)\Tr(\Lambda D_\Lambda)
+:\e_2\Big(\Tr(\Lambda D_\Lambda)^k\Big):\label{B2m}\\
B_3&=&3\e_3\Big(\Tr\Lambda^k\Big)+\Tr(\Lambda^3 D_\Lambda)-\e_1\Big(\Tr\Lambda^k\Big)\Tr(\Lambda^2 D_\Lambda)
+\e_2\Big(\Tr\Lambda^k\Big)\Tr(\Lambda D_\Lambda)+\nn\\
&+&{1\over 2}\e_1\Big(\Tr\Lambda^k\Big):\e_2\Big(\Tr(\Lambda D_\Lambda)^k\Big):+{1\over 2}:\e_3\Big(\Tr(\Lambda D_\Lambda)^k\Big):-\nn\\
&-&{1\over 2}:\Big[\left(\Tr\Lambda^2 D_\Lambda\right)\left(\Tr\Lambda D_\Lambda\right)-\Tr \Lambda^2 D_\Lambda \Lambda D_\Lambda\Big]:\label{B3m}
\ee
and, generally, at $n>0$,
\be\label{mS1}
\hspace{-1cm}
\boxed{
\begin{array}{rcl}
nB_{n+1}&=&n(n+1)\e_{n+1}\Big(\Tr\Lambda^k\Big)+n\sum_{j=1}^{n+1}(-1)^{j+1}\e_{n-j+1}\Big(\Tr\Lambda^k\Big)T_j+
\sum_{j=1}^n(-1)^{j+1}\e_{n-j}\Big(\Tr\Lambda^k\Big)\sum_{a,b=1}^{a+b=j+1}T_{ab}+\\
&&\\
&+&{1\over 2}\sum_{j=1}^{n-1}(-1)^{j+1}\e_{n-j-1}\Big(\Tr\Lambda^k\Big)\sum_{a,b,c=1}^{a+b+c=j+2}T_{abc}
+\ldots =n(n+1)\e_{n+1}\Big(\Tr\Lambda^k\Big)+\sum_{m=1}{n^{\delta_{m,1}}\over (m-1)!}{\cal C}_{n+1,m}\\
\end{array}
}\nn
\ee
where we defined
\be\label{mS2}
\boxed{
{\cal C}_{n+1,m}=\sum_{j=1}^{n-m+2}(-1)^{j+1}
\e_{n-j-m+2}\Big(\Tr\Lambda^k\Big)\sum_{a_1,\ldots,a_m=1}^{\sum_{i=1}^m=j+m-1}T_{a_1,\ldots,a_m}
}
\ee
and
\be
T_a&=&\Tr\Big(\Lambda^a D_\Lambda\Big)\nn\\
T_{ab}:&=&{1\over 2}:\left[\Tr\Big(\Lambda^a D_\Lambda\Big)\Tr\Big(\Lambda^b D_\Lambda\Big)-\Tr\Big(\Lambda^a D_\Lambda
\Lambda^b D_\Lambda\Big)\right]:\nn\\
T_{abc}:&=&:\left[{1\over 6}\Tr\Big(\Lambda^a D_\Lambda\Big)\Tr\Big(\Lambda^b D_\Lambda\Big)\Tr\Big(\Lambda^c D_\Lambda\Big)
-{1\over 2}\Tr\Big(\Lambda^a D_\Lambda\Lambda^b D_\Lambda\Big)\Tr\Big(\Lambda^c D_\Lambda\Big)+{1\over 3}
\Tr\Big(\Lambda^a D_\Lambda\Lambda^b D_\Lambda \Lambda^c D_\Lambda\Big)\right]:\nn\\
&\ldots&
\ee
In terms of the permutation group ${\cal S}_m$, these functions are\footnote{
In other words, one can construct $T_{a_1,\ldots,a_m}$ in the following way. One starts from the elementary symmetric polynomial, and
\be
T_{1,1,\ldots,1}=\e_m\Big(\Tr (\Lambda D_\Lambda)^k\Big)
\ee
This polynomial is a sum of monomials constructed from $m$ entries $\Lambda D_\Lambda$ each. Now, one just replaces each $\Lambda D_\Lambda$ with $\Lambda^{a_i}D_\Lambda$, this is just $T_{a_1,\ldots,a_m}$. In other words, each monomial is of the form
\be
\Lambda_{i_1j_1}\Lambda_{i_2j_2}\ldots\Lambda_{i_mj_m}\Big(D_\Lambda\Big)_{i_1,\sigma(j_1)}\Big(D_\Lambda\Big)_{i_2,\sigma(j_2)}
\ldots \Big(D_\Lambda\Big)_{i_m,\sigma(j_m)}
\ee
where $\sigma$ is a permutation of $m$ elements, and one replaces it with
\be
\Lambda_{i_1j_1}^{a_1}\Lambda_{i_2j_2}^{a_2}\ldots\Lambda_{i_mj_m}^{a_m}\Big(D_\Lambda\Big)_{i_1,\sigma(j_1)}\Big(D_\Lambda\Big)_{i_2,\sigma(j_2)}
\ldots \Big(D_\Lambda\Big)_{i_m,\sigma(j_m)}
\ee
}
\be\label{mS3}
\boxed{
T_{a_1,\ldots,a_m}=:{1\over m!}\sum_\sigma \sgn(\sigma) \Lambda_{i_1j_1}^{a_1}\Lambda_{i_2j_2}^{a_2}\ldots\Lambda_{i_mj_m}^{a_m}\Big(D_\Lambda\Big)_{i_1,\sigma(j_1)}\Big(D_\Lambda\Big)_{i_2,\sigma(j_2)}
\ldots \Big(D_\Lambda\Big)_{i_m,\sigma(j_m)}:
}
\ee
where the sum runs over all permutations $\sigma\in {\cal S}_m$.

Note that the operators ${\cal C}_{n,m}$ can be conveniently encoded in the generating function:
\be
\exp\left(-\sum_{k,m}\Big(\Tr\Lambda^k\Big){(-1)^{k+1}z^k\over k}\right)\times:\exp\left(-\sum_{k,m}\Tr\Big({y\Lambda\over 1+z\Lambda} D_\Lambda\Big)^k {(-1)^{k+1}z^k\over k}\right):=\sum_n{\cal C}_{n,m}z^ny^m
\ee

In the one-body representation (i.e. $N=1$), which is a defining representation at zero central charge, the operators $B_k$ trivialize to
\be
B_1&=&\e_1\Big(\Tr\Lambda^k\Big)+\Tr(\Lambda D_\Lambda)=\Tr\Lambda+\Tr(\Lambda D_\Lambda)\to z(1+D_z)\\
(-1)^nB_n&=&\e_1\Big(\Tr\Lambda^k\Big)\Tr(\Lambda^{n-1} D_\Lambda)-\Tr(\Lambda^n D_\Lambda)=
\Tr\Lambda\cdot\Tr(\Lambda^{n-1} D_\Lambda)-\Tr(\Lambda^n D_\Lambda)\to 0\ \ \ \ \ \ \hbox{at}\ n>1\nn
\ee
and, hence, are not described as elements of the Lie algebra $W_{1+\infty}$.

\section{Relation between matrix and Fock representations}

Expression (\ref{b2})  for $\hat b_2$  at $\beta=1$ is nicely consistent with (\ref{B2m})
\be
2:\e_2\Big(\Tr(\Lambda D_\Lambda)^k\Big):\  = \  :\Big(\Tr(\Lambda D_\Lambda)\Big)^2 - \Tr(\Lambda D_\Lambda)^2:
= \sum_{i,j,k,l=1}^N \Lambda^2_{ij} \Lambda^2_{kl}
\left(\frac{\p}{\p\Lambda_{ij}}\frac{\p}{\p \Lambda_{kl}}-\frac{\p}{\p\Lambda_{kj}}\frac{\p}{\p \Lambda_{il}}\right)
= \nn \\
= \sum_a \left(\sum_{i,j,k,l=1}^N \Lambda^2_{ij} \Lambda^2_{kl}
\left(\frac{\p}{\p\Lambda_{ij}} \Lambda^{a-1}_{lk} - \frac{\p}{\p\Lambda_{kj}} \Lambda^{a-1}_{li} \right)
a\frac{\p}{\p p_a}\right) = \nn \\
= \sum_{a,b=1} ab\Big(\underline{\underline{p_{a+1}p_{b+1} }}
-\underline{p_{a+b+2}}
\Big)\frac{\p^2}{\p p_a\p p_b}
+\underbrace{\sum_{a=1} \left(\sum_{c+d=a-2} \big(p_{c+d+4}-p_{c+2}p_{d+2}  \big)\right) a \frac{\p}{\p p_a}}_{
\underline{\sum_{a=1} a(a-1)p_{a+2}\frac{\p}{\p p_a}} - \underline{\underline{\sum_{a,b=1}(a+b)p_{a+1}p_{b+1}\frac{\p}{\p p_{a+b}}}}
}
= \nn \\
= \boxed{ \sum_{a,b=1}\Big(\underline{\underline{p_{a+1}p_{b+1} }} -\underline{p_{a+b+2}}\Big)\left(ab\frac{\p^2}{\p p_a\p p_b}
- (a+b)\frac{\p}{\p p_{a+b}}\right) }
=  \left.2\hat b_2\right|_{\beta=1}
\ee
where once and double underlined terms are parts of  $2\hat a_2^{(1)}$  and $2\hat a_2^{(2)}$ respectively. Note that the first three terms in (\ref{B2m}) are just the first two terms,
$3\e_2(p_k)  +  [\hat \psi_3, \e_2(p_k)]$ in (\ref{Bmp}).

Similarly, for (\ref{B3m}) and $m=3$,
\be
6:\e_3\Big(\Tr(\Lambda D_\Lambda)^k\Big): \ = \  :\Big(\Tr(\Lambda D_\Lambda)\Big)^3 - 3 \Tr(\Lambda D_\Lambda)^2 \Tr\Lambda D_\Lambda
+ 2\Tr(\Lambda D_\Lambda)^3 : \ =\nn\\
=  \sum_{a,b,c=1}
\Big(p_{a+1}p_{b+1}p_{c+1} - p_{a+1}p_{b+c+2} - 2p_{a+b+2}p_{c+1} +2p_{a+b+c+3}\Big)
\cdot\nn \\ \cdot
\left(abc \frac{\p^3}{\p p_a\p p_b\p p_c} - 3a(b+c) \frac{\p^2}{\p p_a \p p_{b+c}} + 2  (a+b+c)\frac{\p}{\p p_{a+b+c}}\right)
\ee
see Appendix B for the derivation.
This expression annihilates $S_{[r]}$ (moreover, differently underlined terms do this independently).

The complete answer in this case is
\be
\hat B_3 = 6\e_3(p_k) + [\hat\psi_3, \e_3(p_k)]
+\frac{1}{12}\sum_{a,b,c=1} \Big(p_{a+1}p_{b+1}p_{c+1} - p_{a+1}p_{b+c+2} - 2p_{a+b+2}p_{c+1} +2p_{a+b+c+3}\Big)
\cdot\nn \\ \cdot
\underbrace{\left(abc \frac{\p^3}{\p p_a\p p_b\p p_c} - 3a(b+c) \frac{\p^2}{\p p_a \p p_{b+c}}
+ 2  (a+b+c)\frac{\p}{\p p_{a+b+c}}\right)}_{\hat A^{(3)}_{a,b,c}}
+ \nn \\
+\frac{1}{4}\sum_{a,b=1}\Big( p_1\big(p_{a+1}p_{b+1}-p_{a+b+2}\big)-2\big(p_{a+1}p_{b+2}-p_{a+b+3}\big)\Big)
\underbrace{\left(ab\frac{\p^2}{\p p_a\p p_b} - (a+b)\frac{\p}{\p p_{a+b}}\right)}_{\hat A^{(2)}_{ab}}
\ee
Then
\be
\hat B_3 S_{[r_1,r_2,r_3]} = \frac{(r_1+3)(r_2+2)(r_3+1)}{2}S_{[r_1+1,r_2+1,r_3+1]}
\ee

\section{Fock representation: the general answer}

The results of the previous section urge to look for a general operator $\hat B_m$ in the form
\be\label{Bmpg}
\boxed{
\hat B_m = \frac{m(m+1)}{2} \e_m(p_k) + \left[\hat\psi_3, \e_m(p_k)\right] + \sum_{n=2}^m
\left(\sum_{a_1,\ldots,a_n} {\bf K}_{(m,n)}^{a_1,\ldots,a_n}  \hat A^{(n)}_{a_1,\ldots,a_n}\right)
}
\ee
The explicit forms of operators $\hat A^{(n)}$ and polynomials $K_{(m,n)}(p)$ can be read from the matrix expressions.

\paragraph{Operators $\hat A^{(n)}$.}
The  operators $\hat A$ annihilate all $S_{[r]}$ for arbitrary values of $a,b,c$.
The structure of these operators is dictated in an obvious way by the shapes of $2\e_2(p_k)=p_1^2-p_2$ and $3\e_3(p_k)=p_1^3-3p_2p_1+2p_3$.
One can actually choose them symmetric:
\be
\hat A^{(2)}_{a,b} &=& ab\frac{\p^2}{\p p_a\p p_b} - (a+b)\frac{\p}{\p p_{a+b}}, \nn \\
\hat A^{(3)}_{a,b,c} &=& abc \frac{\p^3}{\p p_a\p p_b\p p_c} - a(b+c) \frac{\p^2}{\p p_a \p p_{b+c}}
 - b(a+c) \frac{\p^2}{\p p_b \p p_{a+c}}  - c(a+b) \frac{\p^2}{\p p_c \p p_{a+b}}
+ 2  (a+b+c)\frac{\p}{\p p_{a+b+c}}
\ee

The next such operators, induced by $24\e_4(p_k) = p_1^4-6p_2p_1^2+8p_3p_1+3p_1^2-6p_4$
and $120\e_5(p_k) = p_1^5-10p_2p_1^3 +20p_3p_1^2+15p_2^2p_1-30p_4p_1-20p_3p_2+24p_5$ are
\be
\hat A^{(4)}_{a,b,c,d} ={\rm Sym}_{a,b,c,d} \left\{abcd\frac{\p^4}{\p p_a\p p_b\p p_c\p p_d} - 6ab(c+d)\frac{\p^3}{\p p_a\p p_b\p p_{c+d}}
+ 8 a(b+c+d)\frac{\p^2}{\p p_a\p p_{b+c+d}}
+ \right. \nn \\ \left.
+3(a+b)(c+d)\frac{\p^2}{\p p_{a+b}\p p_{c+d}}
- 6(a+b+c+d)\frac{\p}{\p p_{a+b+c+d}}\right\}
\ee
and
\be
\hat A^{(5)}_{a,b,c,d,e} ={\rm Sym}_{a,b,c,d,e} \left\{abcde\frac{\p^5}{\p p_a\p p_b\p p_c\p p_d\p p_e}
-10 abc(d+e)\frac{\p^4}{\p p_a\p p_b\p p_c\p p_{d+e}}
+ \right.\nn \\ \left.
+20ab(c+d+e)\frac{\p^3}{\p p_a\p p_b\p p_{c+d+e}} + 15a(b+c)(d+e)\frac{\p^3}{\p p_a\p p_{b+c}\p p_{d+e}}
- \right.\nn \\ \left.
-30a(b+c+d+e)\frac{\p^2}{\p p_a\p p_{b+c+d+e}} - 20(a+b)(c+d+e)\frac{\p^2}{\p p_{a+b}\p p_{c+d+e}}
+ \right.\nn \\ \left.
 + 24(a+b+c+d+e)\frac{\p}{\p p_{a+b+c+d+e}}
\right\}
\ee
They also annihilate all $S_{[r]}$. Symmetrization here is implied just summing over all permutations without numerical factors.

Generalization of $\hat A^{(n)}$ to higher $n$ is obvious: they are immediately read from $n! \e_n(p_k)$, and have the form
\be\label{AF}
\boxed{
\hat A^{(n)}_{\{a_i\}}=\sum_{\sigma\in{\cal S}_n} \sgn(\sigma)\prod_j \alpha_j
{\p\over\p p_{\alpha_j}}
}
\ee
where the permutation $\sigma$ is realized by a set of cycles $\sigma_j$ in disjoint cycle notation, and $n$ indices are parted into groups associated with each cycle: $\alpha_j=\sum_{k=1}^{\sigma_j} a_{k,j}$.

\paragraph{Polynomials $K_{(m,n)}(p)$.} It remains to identify the polynomials $K_{(m,n)}(p)$ in (\ref{Bmpg}). There
\be
{\bf K}_{(2,2)}^{ab} = \frac{1}{2} p_{a+1}p_{b+1}-\frac{1}{2}p_{a+b+2} ,
\ee
\be
{\bf K}_{(3,3)}^{abc} &=& \frac{1}{12}\cdot \Big(p_{a+1}p_{b+1}p_{c+1}-p_{a+1}p_{b+c+2}-p_{b+1} p_{a+c+2}
- p_{c+1} p_{a+b+2}  +2p_{a+b+c+3}\Big),
\nn \\
{\bf K}_{(3,2)}^{ab} &=& p_1{\bf K}_{(2,2)}^{ab} - \frac{1}{4}\Big(p_{a+1}p_{b+2}-p_{a+2}p_{b+1}\Big)+\frac{1}{2} p_{a+b+3}\Big)
\ee
with a symmetrized choice of $\hat A_3$ one can use shorter, asymmetric versions:
\be
{\bf K}_{(3,3)}^{abc}& =& \frac{1}{12}\cdot \Big(p_{a+1}p_{b+1}p_{c+1}-3p_{a+1}p_{b+c+2}\Big) +\frac{1}{6}p_{a+b+c+3},
\nn \\
{\bf K}_{(3,2)}^{ab} &= &\frac{1}{2} p_1{\bf K}_{(2,2)}^{ab} - \frac{1}{2} p_{a+1}p_{b+2}+ \frac{1}{2}p_{a+b+3} ,
\ee
\be
{\bf K}_{(4,4)}^{abcd}& =& \frac{1}{144}\Big( p_{a+1}p_{b+1}p_{c+1}p_{d+1} - 6p_{a+b+2}p_{c+1}p_{d+1}
+ 8p_{a+b+c+3}p_{d+1} + 3p_{a+b+2}p_{c+d+2} - 6p_{a+b+c+d+4}\Big),
\nn \\
{\bf K}_{(4,3)}^{abc} &=& \frac{1}{3}p_1{\bf K}_{(3,3)}^{abc}
- \frac{1}{36}\Big(\big(p_{a+2}p_{b+1}p_{c+1} + p_{a+1}p_{b+2}p_{c+1} + p_{a+1}p_{b+1}p_{c+2}\big) -  \nn \\
&-&2\big(p_{a+1}p_{b+c+3}+p_{b+1}p_{a+c+3}+p_{c+1}p_{a+b+3}\big) - \big(p_{a+2}p_{b+c+2}+p_{b+2}p_{a+c+2}+p_{c+2}p_{a+b+2}\big)
+ 6 p_{a+b+c+4}\Big),
\nn\\
{\bf K}_{(4,2)}^{ab} &=& \frac{1}{3}\e_2(p_k) {\bf K}_{(2,2)}^{ab} - \frac{1}{6} p_1(p_{a+1}p_{b+2}+p_{a+2}p_{b+1}-2p_{a+b+3})
+ \frac{1}{6}\Big(p_{a+3}p_{b+1} + p_{a+2}p_{b+2} + p_{a+1}p_{b+3}\Big) - \frac{1}{2}p_{a+b+4}\nn\\
\ee
abbreviated non-symmetric version to be used with the symmetrized $\hat A$ being
\be
{\bf K}_{(4,3)}^{abc} &=& \frac{1}{3}p_1{\bf K}_{(3,3)}^{abc}
- \frac{1}{12}\Big( p_{a+2}p_{b+1}p_{c+1}
-2 p_{a+1}p_{b+c+3}  - p_{a+2}p_{b+c+2}\Big)
-\frac{1}{6} p_{a+b+c+4},
\nn\\
{\bf K}_{(4,2)}^{ab} &=&\frac{1}{3}\e_2(p_k) {\bf K}_{(2,2)}^{ab} - \frac{1}{3} p_1(p_{a+1}p_{b+2}-p_{a+b+3})
+ \frac{1}{6} \Big(2p_{a+3}p_{b+1} +p_{a+2}p_{b+2}\Big) -  \frac{1}{2}p_{a+b+4}
\ee
If one introduce the notation
\be
{\bf K}_n^{(a)}(a_1,a_2,\ldots,a_n)=\sum_{c_1,c_2,\ldots,c_n=0}^{\sum_kc_k=a}{\bf K}_{(n,n)}(a_1+c_1,a_2+c_2,\ldots,a_n+c_n)
\ee
these formulas in the symmetric version can be rewritten in an inspiring form
\be
{\bf K}_{(3,2)} &= &{1\over 2}\e_1(p_k){\bf K}_2^{(0)}-{1\over 2}{\bf K}_2^{(1)}\nn\\
{\bf K}_{(4,3)} &= &{1\over 3}\e_1(p_k){\bf K}_3^{(0)}-{1\over 3}{\bf K}_3^{(1)}\nn\\
{\bf K}_{(4,2)} &= &{1\over 3}\e_2(p_k){\bf K}_2^{(0)}-{1\over 3}\e_1(p_k){\bf K}_2^{(1)}+{1\over 3}{\bf K}_2^{(2)}
\ee

The generic expression for polynomials ${\bf K}_{(n,m)}$ is now also rather evident and looks much similar to formulas of sec.\ref{mr}.

As in sec.\ref{mr}, we first construct the polynomial ${\bf K}_{(n,n)}$:
\be\label{K1F}
\boxed{
{\bf K}_{(n,n)}={1\over (n-1)!n!}\sum_{\sigma\in{\cal S}_n} \sgn(\sigma)\prod_j p_{\bar\alpha_j}
}
\ee
where $\bar\alpha_j:=\sum_{k=1}^{\sigma_j} (a_{k,j}+1)$. Now the construction of an arbitrary ${\bf K}_{(n,m)}$ is immediate and much similar to formula (\ref{mS2}):
\be\label{K2F}
\boxed{
{\bf K}_{(n,m)}(a_1,a_2,\ldots,a_m)=\binom{n-1}{m-1}^{-1}\cdot
\sum_{j=0}^{n-m} (-1)^j\e_{n-m-j}(p_k)
\underbrace{\sum_{c_1,c_2,\ldots,c_m=0}^{\sum_kc_k=j}{\bf K}_{(m,m)}(a_1+c_1,a_2+c_2,\ldots,a_m+c_m)}_{{\bf K}_m^{(j)}}
}
\ee

More examples of polynomials ${\bf K}_{(n,m)}$ can be found in Appendix C.

\section{Fock representation of affine Yangian algebra: Jack polynomials}

In the case of the Jack polynomials, which are associated with the Fock representation of the affine Yangian algebra $Y(\widehat{gl}_1)$ the creation operators $\hat B_m$ are modified as follows:
\be\label{BmpgB}
\boxed{
\hat B^{(\beta)}_m = \frac{m(m+1)}{2}\beta^m \e_m(p_k) + \beta^{m-1} \left[\hat\psi_3^{(\beta)}, \e_m(p_k)\right] + \sum_{n=2}^m \beta^{m-n}
\left(\sum_{a_1,\ldots,a_n} {\bf K}_{(m,n)}^{a_1,\ldots,a_n}  \hat A^{(\beta,n)}_{a_1,\ldots,a_n} \right)
}
\ee
As compared with formulas of the previous section, this is just simple rescalings and modifications of the operators $\hat A^{(n)}$ and $\hat\psi_3^{\beta)}$. The latter one depends on $\beta$ according to (\ref{psi23}),
this is a little more than rescaling, one can call it anomaly.
The polynomials $K_{(m,n)}(p)$ are $\beta$-independent and stay the same.

The operators $\hat A^{(n)}$ look now as
\be
\hat A^{(\beta,2)}_{a,b} &=& ab\frac{\p^2}{\p p_a\p p_b} - \beta(a+b)\frac{\p}{\p p_{a+b}},
\nn\\
\hat A^{(\beta,3)}_{a,b,c} &=& abc \frac{\p^3}{\p p_a\p p_b\p p_c} - \beta\left( a(b+c) \frac{\p^2}{\p p_a \p p_{b+c}}
 + b(a+c) \frac{\p^2}{\p p_b \p p_{a+c}}  + c(a+b) \frac{\p^2}{\p p_c \p p_{a+b}}\right)+\nn\\
&+& 2\beta^2  (a+b+c)\frac{\p}{\p p_{a+b+c}}, \nn\\
\hat A^{(\beta,4)}_{a,b,c,d} &=&{\rm Sym}_{a,b,c,d} \left\{abcd\frac{\p^4}{\p p_a\p p_b\p p_c\p p_d} - 6\beta ab(c+d)\frac{\p^3}{\p p_a\p p_b\p p_{c+d}}
+ 8\beta a(b+c+d)\frac{\p^2}{\p p_a\p p_{b+c+d}}
+ \right. \nn \\
&+&\left. 3\beta^2 (a+b)(c+d)\frac{\p^2}{\p p_{a+b}\p p_{c+d}}
- 6 \beta^3(a+b+c+d)\frac{\p}{\p p_{a+b+c+d}}\right\},
\nn\\
\hat A^{(\beta,5)}_{a,b,c,d,e} &=&{\rm Sym}_{a,b,c,d,e} \left\{abcde\frac{\p^5}{\p p_a\p p_b\p p_c\p p_d\p p_e}
-10\beta abc(d+e)\frac{\p^4}{\p p_a\p p_b\p p_c\p p_{d+e}}
+ \right.\nn \\
&+&20\beta^2 ab(c+d+e)\frac{\p^3}{\p p_a\p p_b\p p_{c+d+e}} + 15\beta^2 a(b+c)(d+e)\frac{\p^3}{\p p_a\p p_{b+c}\p p_{d+e}}
- \nn \\
&-&30\beta^3 a(b+c+d+e)\frac{\p^2}{\p p_a\p p_{b+c+d+e}} - 20\beta^3 (a+b)(c+d+e)\frac{\p^2}{\p p_{a+b}\p p_{c+d+e}}
+ \nn \\
 &+& \left. 24\beta^4(a+b+c+d+e)\frac{\p}{\p p_{a+b+c+d+e}}
\right\}
\ee
and, generally,
\be\label{BAF}
\boxed{
\hat A^{(\beta,n)}_{\{a_i\}}=
\beta^{n-1} \sum_{\sigma\in{\cal S}_n} \sgn(\sigma)\prod_j \alpha_j
{\p\over\beta \p p_{\alpha_j}}
}
\ee

Operators $\hat B^{(\beta)}_m$ act on Jack polynomials by the rule:
\be
\boxed{
\hat B^{(\beta)}_m J_{[r_1r_2\ldots r_m]} = \frac{(r_1+m\beta)(r_2+(m-1)\beta)\ldots(r_m+\beta)}{(m-1)!}J_{[r_1+1,r_2+1,\ldots r_m+1]}
}
\label{BJack}
\ee
Normalization of $\hat B$ can be chosen differently, at present
\be
\hat B^{(\beta)}_m \cdot 1 = \frac{m(m+1)}{2}\beta^m \e_m(p_k)   + \beta^{m-1} \left[\hat\psi_3^{(\beta)}, \e_m(p_k)\right]
= m \beta^m \e_m(p_k)
\ee

\section{Conclusion}

In this paper, we have constructed the operators $\hat B_m$ adding columns to Young diagrams in the matrix representation of the $W_{1+\infty}$ algebra (Eqs. (\ref{mS1}), (\ref{mS2}), (\ref{mS3})), in the Fock representation of $W_{1+\infty}$ algebra (Eqs. (\ref{Bmpg}), (\ref{AF}), (\ref{K1F}), (\ref{K2F})), and in the Fock representation of the affine Yangian algebra $Y(\widehat{gl}_1)$ (Eqs. (\ref{BmpgB}), (\ref{BAF}), (\ref{K1F}), (\ref{K2F})).

We find the idea of creation operators for symmetric polynomials very interesting
and deserving more attention than it received so far.
The operators are unusual, because they are not commuting and add lines to Young diagrams
in a fixed order.
It may happen that such relaxation of the standard requirement of commutativity will be fruitful in a much broader context.

Despite the polynomials from the Schur-Jack-Macdonald family are related to the Gaussian Hermitian matrix models and its extensions,
the Gaussianity is long known to interfere non-trivially with non-Abelianity:
we know this in examples of free-field representations in conformal models \cite{DF,Wak,FF,GMMOS},
in Chern-Simons theory \cite{CS,Gua,MS} and in the recent far-going Yangian and DIM generalizations/applications
of the matrix models themselves \cite{Ch2,Ch3,MMP}.
Perhaps, the formalism of non-commuting creation operators will help in understanding the non-Abelian
phenomena in Gaussian matrix models and their remarkable (though hidden) superintegrability properties \cite{SIrev,SIU,MMNek}.

Moreover, the new rich  system of WLZZ models \cite{China1,China2,MMsc,Ch1,Ch2,Ch3} demonstrates a possibility of using the same symmetric
polynomials far beyond Gaussian models, preserving and even broadening their basic properties,
from $W$-representations to highly extended multi-integrability and superintegrability.
It looks challenging to unify these essentially non-linear properties with existence of the simple, even non-commutative
creation operators, which implies existence of an additional, yet uncovered hierarchical structure.

\bigskip

There are some other straightforward questions, left open in the present paper:

\begin{itemize}
\item The system of creation operators is of course complemented by that of annihilation ones.
It does not bring in too many new ideas, but the exposition would double the volume of the text,
thus we decided not to include it.
Still, it is an interesting direction, absolutely necessary for the full description of Pieri rules
and their use in construction of representation theory in terms of Young diagrams, in particular, along the lines of
\cite{AT}.

\item A puzzling part of the story concerns relation to the theory of orthogonal polynomials.
This is another view on the Schur-Macdonald family, with further generalization to Kerov polynomials \cite{Kerov,MMkerov}.
However, non-commutativity and sophisticated hierarchy of creation operators make relation to orthogonality
a little puzzling.

\item Generalization of \cite{KN} to the Fock representation should be also straightforward for other polynomials beyond the Schur and Jack ones. Primarily this should be true for the Macdonald polynomials.
Also interesting should be generalizations to the Macdonald polynomials associated with all root systems \cite{Macrs,Chrs},
which can also raise questions concerning the Vogel's universality \cite{Vogel,BM,BMM,MSl,B}.

\item The original Kirillov-Noumi problem was not restricted to symmetric polynomials,
their creation/annihilation operators can act also in the space of non-symmetric one.
This direction is especially interesting, because an algebraic interpretation and representation theory
here is much less studied.
Technically, there are no eigenvalue matrix models, no power sum variables, no clear notion of superintegrability in the non-symmetric case.
Thus, the operator formalism, once developed, can play a role of the substitute of all their standard machinery,
which can be useful both for the new field of applications and for a deeper understanding of the basic theory itself.
\end{itemize}

We expect that all these issues attract attention in the near future, and hope to return to them elsewhere.

\section*{Acknowledgements}

We are grateful to I. Ryzhkov for a useful discussion. The work was partially funded within the state assignment of the Institute for Information Transmission Problems of RAS. Our work is also partly supported by the grant of the Foundation for the Advancement of
Theoretical Physics and Mathematics “BASIS”.

\section*{Appendix A}

In this Appendix, as a continuation of sec.2, we reproduce an explicit construction of the operators $\hat B_m$ in the $N$-body representation in the case of Macdonald polynomials due to \cite{KN}. In fact, there are two sets of formulas. One of them is based on the Cherednik operators as a deformation of the Dunkl operators and literally extends formula (\ref{KNops}) to the Macdonald case:
\be
\hat B_m = \sum_{1\le k_1<\ldots<k_m\le N}  x_{k_1}\ldots x_{k_m}(1-t\,Ch_{k_1})(1-t^2Ch_{k_2})\ldots (1-t^mCh_{k_m})
\ee
where $Ch_k$ is the Cherednik operator \cite{Ch,paper:NS-cherednik}
\begin{align}
    Ch_k = t^{1 - k} \left(\prod_{i = k + 1}^N R_{k,i}\right)
    \ttop{k}
    \left(\prod_{i = 1}^{k - 1} R^{-1}_{i,k}\right)\nn
  \end{align}
with the $R$-operators equal to
  \begin{align}
    R_{i,j} = \frac{(x_i - t^{-1} x_j)}{(x_i - x_j)}
    + \left(t^{-1} - 1\right)\frac{x_j}{(x_i - x_j)} \hat\sigma_{i,j}\nn
    \\ \notag
    R^{-1}_{i,j} =
    t \frac{(t^{-1} x_i - x_j)}{(x_i - x_j)}
    + \left(t - 1 \right)\frac{x_j}{(x_i - x_j)} \hat\sigma_{i,j}
  \end{align}
In particular, one has
  \begin{align}
    R^{-1}_{i,j} R_{i,j} = & \ \text{Id}\nn
  \end{align}
Cherednik operators for $N$ particles naturally emerge within the framework of the $GL_N$ affine Hecke algebra, see a manifest description in \cite{Ch}) (see also \cite[Appendix B]{MMP}).

Another realization of the operators $\hat B_m$ is available \cite[Formula (6)]{KN} when they act on symmetric functions (which, in fact, we are interested in):
\be
\hat B_m =\sum_{r=0}^m(-1)^rt^{{r(r-1)\over 2}+(m-N+1)r}\sum_{|I|=r, I\in {\cal N}}
\e_{m-r}[x_{k\in {\cal N}/I}]\cdot {\left(\prod_{j\in I}t^{D_j}\Delta(x)\right)\over \Delta(x)}
\cdot\prod_{j\in I}x_jq^{D_j}
\ee
where ${\cal N}$ denotes the set of $N$ first natural numbers, $D_j:=x_j{\p\over\p x_j}$, and $\Delta(x)=\prod_{i<j}(x_i-x_j)$ is the Vandermonde determinant.

On the lines of the present paper interesting  is the lift of this formulas to Fock representation,
with the simplest example provided by Eq.(\ref{MacB1}).
Full construction in the Fock representation, generalizing (\ref{BJack}) to the Macdonald level will be presented elsewhere.

\section*{Appendix B}

\be
6:\e_3\Big(\Tr(\Lambda D_\Lambda)^k\Big): \ = \  :\Big(\Tr(\Lambda D_\Lambda)\Big)^3 - 3 \Tr(\Lambda D_\Lambda)^2 \Tr\Lambda D_\Lambda
+ 2\Tr(\Lambda D_\Lambda)^3 : \
= \nn \\
= \sum_{i,j,k,l,m.n=1}^N \Lambda^2_{ij} \Lambda^2_{kl}  \Lambda^2_{mn}
\left(\frac{\p}{\p\Lambda_{ij}}\frac{\p}{\p \Lambda_{kl}}\frac{\p}{\p \Lambda_{mn}}
-3\frac{\p}{\p\Lambda_{kj}}\frac{\p}{\p \Lambda_{il}}\frac{\p}{\p \Lambda_{mn}}
+ 2 \frac{\p}{\p\Lambda_{kj}}\frac{\p}{\p \Lambda_{ml}}\frac{\p}{\p \Lambda_{in}}\right)
= \nn \\
= \sum_a \left(\sum_{i,j,k,l,m.n=1}^N \Lambda^2_{ij} \Lambda^2_{kl}  \Lambda^2_{mn}
\left(\frac{\p}{\p\Lambda_{ij}}\frac{\p}{\p \Lambda_{kl}}  \Lambda^{a-1}_{nm}
-3\frac{\p}{\p\Lambda_{kj}}\frac{\p}{\p \Lambda_{il}}\Lambda^{a-1}_{nm}
+ 2 \frac{\p}{\p\Lambda_{kj}}\frac{\p}{\p \Lambda_{ml}}\Lambda^{a-1}_{ni}\right)
a \frac{\p}{\p p_a}\right)
= \nn \\
\nn \\
=  \sum_{i,j,k,l,m.n=1}^N \Lambda^2_{ij} \Lambda^2_{kl}  \Lambda^2_{mn}
\left\{
\sum_{a,b=1} \left(\frac{\p}{\p\Lambda_{ij}}  \Lambda^{b-1}_{lk}  \Lambda^{a-1}_{nm}
-3\frac{\p}{\p\Lambda_{kj}}\Lambda^{b-1}_{li}\Lambda^{a-1}_{nm}
+ 2 \frac{\p}{\p\Lambda_{kj}}\Lambda^{b-1}_{lm}\Lambda^{a-1}_{ni}\right)
ab \frac{\p^2}{\p p_a\p p_b}
+ \right. \nn
\ee
\be
\left.
+ \sum_{b,c=0} \left(\frac{\p}{\p\Lambda_{ij}}   \Lambda^{b}_{nk} \Lambda^{c}_{lm}
-3\frac{\p}{\p\Lambda_{kj}} \Lambda^{b}_{ni} \Lambda^{c}_{lm}
+ 2 \frac{\p}{\p\Lambda_{kj}} \Lambda^{b}_{nm} \Lambda^{c}_{li}\right)
(b+c+2) \frac{\p}{\p p_{b+c+2}} \right\}
=
\nn\\
= \sum_{i,j,k,l,m.n=1}^N \Lambda^2_{ij} \Lambda^2_{kl}  \Lambda^2_{mn}
\left\{
\sum_{a,b,c=1} \left(\Lambda^{c-1}_{ji}  \Lambda^{b-1}_{lk}  \Lambda^{a-1}_{nm}
-3\Lambda^{c-1}_{jk}\Lambda^{b-1}_{li}\Lambda^{a-1}_{nm}
+ 2 \Lambda^{c-1}_{jk}\Lambda^{b-1}_{lm}\Lambda^{a-1}_{ni}\right)
abc \frac{\p^3}{\p p_a\p p_b\p p_c}
+ \right. \nn \\ \left.
+\sum_{a =1}\sum_{c,d=0} \left(  \Lambda^{c}_{li}\Lambda^{d}_{jk}  \Lambda^{a-1}_{nm}
-3  \Lambda^{c}_{lk}\Lambda^{d}_{ji}\Lambda^{a-1}_{nm}
+ 2   \Lambda^{c}_{lk}\Lambda^{d}_{jm}\Lambda^{a-1}_{ni}\right)
a(c+d+2) \frac{\p^2}{\p p_a\p p_{c+d+2}}
+ \right. \nn \\ \left.
 +\sum_{b=1}\sum_{c,d=0} \left(   \Lambda^{b-1}_{lk}  \Lambda^{c}_{ni}\Lambda^{d}_{jm}
-3 \Lambda^{b-1}_{li}\Lambda^{c}_{nk}\Lambda^{d}_{jm}
+ 2 \Lambda^{b-1}_{lm}\Lambda^{c}_{nk}\Lambda^{d}_{ji}\right)
(c+d+2)b \frac{\p^2}{\p p_{c+d+2}\p p_b}
+ \right. \nn \\ \nn \\ \left.
+ \sum_{a=1}\sum_{b,c=0} \left(  \Lambda^{a-1}_{ji}   \Lambda^{b}_{nk} \Lambda^{c}_{lm}
-3   \Lambda^{a-1}_{jk} \Lambda^{b}_{ni} \Lambda^{c}_{lm}
+ 2   \Lambda^{a-1}_{jk} \Lambda^{b}_{nm} \Lambda^{c}_{li}\right)
a(b+c+2) \frac{\p^2}{\p p_a \p p_{b+c+2}}
+ \right. \nn \\ \nn \\ \left.
+  \sum_{c,d,e=0} \left(   \Lambda^{d}_{ni}\Lambda^{e}_{jk} \Lambda^{c}_{lm}
-3  \Lambda^{d}_{nk}\Lambda^{e}_{ji} \Lambda^{c}_{lm}
+ 2  \Lambda^{d}_{nk}\Lambda^{e}_{jm} \Lambda^{c}_{li}\right)
(d+e+c+3) \frac{\p}{\p p_{d+e+c+3}}
+ \right. \nn \\  \left.
+   \sum_{b,d,e=0} \left(    \Lambda^{b}_{nk}  \Lambda^{d}_{li}\Lambda^{e}_{jm}
-3  \Lambda^{b}_{ni}  \Lambda^{d}_{lk}\Lambda^{e}_{jm}
+ 2  \Lambda^{b}_{nm}  \Lambda^{d}_{lk}\Lambda^{e}_{ji}\right)
(b+d+e+3) \frac{\p}{\p p_{b+d+e+3}}
\right\}
= \nn \\
= \sum_{a,b,c=1} abc\Big( p_{a+1}p_{b+}p_{c+1} -3p_{a+1}p_{b+c+2}    + 2 p_{a+b+c+3}     \Big) \frac{\p^3}{\p p_a\p p_b\p p_c}
+ \nn \\
+ \sum_{a=1}\sum_{c,d=0} a(c+d+2)\Big(p_{a+1} p_{c+d+4} -3p_{a+1}p_{c+2}p_{d+2} + 2 p_{a+c+3}p_{d+2}\Big)\frac{\p^2}{\p p_a\p p_{c+d+2}}
+ \nn \\
+ \sum_{b=1}\sum_{c,d=0} b(c+d+2)\Big(p_{b+1} p_{c+d+4} -3p_{b+c+d+5}  +2 p_{b+c+3}p_{d+2} \Big)\frac{\p^2}{\p p_b\p p_{c+d+2}}
+ \nn \\
+  \sum_{a=1}\sum_{b,c=0} a(b+c+2)\Big(p_{a+1} p_{b+c+4} -3p_{a+b+c+5}  +2 p_{a+b+3}p_{c+2} \Big)\frac{\p^2}{\p p_a\p p_{b+c+2}}
+ \nn \\
+ \sum_{a,b,c=1} (a+b+c)\Big(4p_{a+b+c+3} -6p_{a+1} p_{b+c+2} + 2p_{a+1}p_{b+1}p_{c+1}\Big)\frac{\p}{\p p_{a+b+c}}
= \nn \\
\nn \\
= \sum_{a,b,c=1}\left\{ abc\Big( p_{a+1}p_{b+}p_{c+1} -3p_{a+1}p_{b+c+2}    + 2 p_{a+b+c+3}     \Big) \frac{\p^3}{\p p_a\p p_b\p p_c}
- \right. \nn \\ \left.
- 3a(b+c)\Big(p_{a+1}p_{b+1}p_{c+1} - p_{a+1}p_{b+c+2}-2p_{b+1}p_{a+c+2} +2p_{a+b+c+3}\Big) \frac{\p^2}{\p p_a \p p_{b+c}}
+ \right. \nn \\ \left.
+2 (a+b+c)\Big(p_{a+1}p_{b+1}p_{c+1}-3p_{a+1}p_{b+c+2}+2p_{a+b+c+3}\Big) \frac{\p}{\p p_{a+b+c}}
\right\}
\nn \\
=  \sum_{a,b,c=1}
\Big(p_{a+1}p_{b+1}p_{c+1} - p_{a+1}p_{b+c+2} - 2p_{a+b+2}p_{c+1} +2p_{a+b+c+3}\Big)
\cdot\nn \\ \cdot
\left(abc \frac{\p^3}{\p p_a\p p_b\p p_c} - 3a(b+c) \frac{\p^2}{\p p_a \p p_{b+c}} + 2  (a+b+c)\frac{\p}{\p p_{a+b+c}}\right)
\ee

\section*{Appendix C}

At level 5, the non-symmetric expressions for the polynomials ${\bf K}_{m,n}$ look as follows:
\be
{\bf K}_{(5,5)}^{abcde} &=& \frac{1}{5!4!}\Big(p_{a+1}p_{b+1}p_{c+1}p_{d+1}p_{e+1} - 10p_{a+1}p_{b+1}p_{c+1}p_{d+e+2}
+ 20p_{a+1}p_{b+1}p_{c+d+e+3}
+ \nn \\
&+&15p_{a+1}p_{b+c+2}p_{d+e+2}-30p_{a+1}p_{b+c+d+e+4} -20p_{a+b+2}p_{c+d+e+3}+24p_{a+b+c+d+e+5}\Big),
\nn \\
{\bf K}_{(5,4)}^{abcd} &=& \frac{1}{4}p_1{\bf K}^{(4,4)}_{abcd} -\frac{1}{144}p_{a+2}p_{b+1}p_{c+1}p_{d+1}
+ \frac{1}{48}\Big(p_{a+b+3}p_{c+1}p_{d+1}+p_{a+b+2}p_{c+2}p_{d+1}\Big)
-\nn \\
&-&\frac{1}{48}p_{a+b+3}p_{c+d+2}-\frac{1}{24}p_{a+b+c+4}p_{d+1}-\frac{1}{72}p_{a+b+c+3}p_{d+2}
+ \frac{1}{24}p_{a+b+c+d+5},
\nn \\
{\bf K}_{(5,3)}^{abc} &=& \frac{1}{6}\Big(3p_1 {\bf K}_{(4,3)}^{abc} -   S_{[2]} {\bf K}_{(3,3)}^{abc} \Big)
- \frac{p_1}{24}\Big( p_{a+2}p_{b+1}p_{c+1}
-2 p_{a+1}p_{b+c+3}  - p_{a+2}p_{b+c+2}\Big)-\nn\\
&-&\frac{p_1}{12} p_{a+b+c+4}+\frac{1}{24} \Big(p_{a+3}p_{b+1}p_{c+1}+ p_{a+2}p_{b+2}p_{c+1}
- 3p_{a+1}p_{b+c+4}-\nn\\
&-&2p_{a+2}p_{b+c+3}-p_{a+3}p_{b+c+2}\Big)
+\frac{1}{6} p_{a+b+c+5},
\nn \\
{\bf K}_{(5,2)}^{ab} &=& \frac{1}{4}\Big(\e_3(p_k) {\bf K}_{(2,2)}^{ab}-\e_2(p_k) ( p_{a+1}p_{b+2}+ p_{a+b+3})
+{1\over 2}p_1(2p_{a+3}p_{b+1} +p_{a+2}p_{b+2} -  3p_{a+b+4})\Big)
-\nn \\
&-& \frac{1}{8} \Big(p_{a+4}p_{b+1} +p_{a+3}p_{b+2}+p_{a+2}p_{b+3}+p_{a+1}p_{b+4}\Big) +\frac{1}{2} p_{a+b+5}
\nn \\
\ldots
\ee
The symmetric version of these formulas can be rewritten as
\be
{\bf K}_{(5,4)} &= &{1\over 4}\e_1(p_k){\bf K}_4^{(0)}-{1\over 4}{\bf K}_4^{(1)}\nn\\
{\bf K}_{(5,3)} &= &{1\over 6}\e_2(p_k){\bf K}_3^{(0)}-{1\over 6}\e_1(p_k){\bf K}_3^{(1)}+{1\over 6}{\bf K}_3^{(2)}\nn\\
{\bf K}_{(5,2)} &= &{1\over 4}\e_3(p_k){\bf K}_2^{(0)}-{1\over 4}\e_2(p_k){\bf K}_2^{(1)}+{1\over 4}\e_1(p_k){\bf K}_2^{(2)}-
{1\over 4}{\bf K}_2^{(3)}
\ee
In particular,
\be
{\bf K}_{(m,2)}^{ab} = \frac{1}{m-1}\sum_{k=2}^{m-1} (1-k) \e_{m-k}(-p_k){\bf K}_{(k,2)}^{ab}
+ \frac{(-)^m}{2(m-1)} \sum_{k=1}^{m-1} p_{a+m-k}p_{b+k}  -\frac{(-)^m}{2} p_{a+b+m}
\ee


\begin{thebibliography}{12}

\bibitem{Mac} I.G. Macdonald, {\it Symmetric functions and Hall polynomials},   Oxford University Press, 1995

\bibitem{UFN3} A. Morozov,
Phys.Usp.(UFN) {\bf 37} (1994) 1,  hep-th/9303139;
hep-th/9502091; hep-th/0502010;\\
A. Mironov, Int.J.Mod.Phys. {\bf A9} (1994) 4355; Phys.Part.Nucl.
{\bf 33} (2002) 537, hep-th/9409190

\bibitem{SIrev} A.~Mironov, A.~Morozov,
Phys.Lett. {\bf B835} (2022) 137573,
arXiv:2201.12917

\bibitem{SIU} A.~Mironov, A.~Morozov, Z.~Zakirova,
Phys. Lett.  \textbf{B831} (2022) 137178,
arXiv:2203.03869

\bibitem{MMNek} A. Mironov, A. Morozov,
Phys. Rev. \textbf{D106} (2022) 126004,
arXiv:2207.08242

\bibitem{schurchar}  R. Stanley, {\it Enumerative Combinatorics}, vol. 2, Cambridge University
Press, New York/Cambridge, 2023.

\bibitem{Cal} F. Calogero, J.Math.Phys. {\bf 12} (1971) 419

\bibitem{Suth} B. Sutherland, Phys.Rev. {\bf A5} (1972) 1372

\bibitem{OP} M. Olshanetsky, A. Perelomov, Phys.Peps. {\bf 71} (1981) 313; Phys.Rep. {\bf 94} (1983) 6

\bibitem{Jack}  H. Jack, Proc. R. Soc. Edinburgh (A) {\bf 69} (1970) 1; (1972) 347

\bibitem{St} R.P. Stanley, Adv. Math. {\bf 77} (1988) 76

\bibitem{Turb} W. Ruehl, A. Turbiner, Mod. Phys. Lett. {\bf A10} (1995) 2213, hep-th/9506105

\bibitem{GP} N. Gurappa, Prasanta K. Panigrahi,
Phys. Rev. {\bf B62} (2000) 1943, hep-th/9910123

\bibitem{Ch1}  A.~Mironov, V.~Mishnyakov, A.~Morozov, A.~Popolitov, R.~Wang, W.Z.~Zhao,
Eur. Phys. J. \textbf{C83} (2023) 377,
arXiv:2301.04107

\bibitem{Ch2} A.~Mironov, V.~Mishnyakov, A.~Morozov, A.~Popolitov, W.Z.~Zhao,
Phys. Lett. \textbf{B839} (2023) 137805,
arXiv:2301.11877

\bibitem{Ch3} Fan Liu, A. Mironov, V. Mishnyakov, A. Morozov, A. Popolitov, Rui Wang, Wei-Zhong Zhao,
  Nucl.Phys. {\bf B993} (2023) 116283,
  arXiv:2303.00552

  \bibitem{MMMP1}  A.~Mironov, V.~Mishnyakov, A.~Morozov, A.~Popolitov,
JHEP \textbf{23} (2020) 065,
arXiv:2306.06623

\bibitem{MMMP2} A.~Mironov, V.~Mishnyakov, A.~Morozov, A.~Popolitov,
Phys. Lett. \textbf{B845} (2023) 138122,
arXiv:2307.01048

\bibitem{MMP} A. Mironov, A. Morozov, A. Popolitov, JHEP \textbf{09} (2024) 200,
  arXiv:2406.16688

\bibitem{Pope} C.N. Pope, L.J. Romans, X. Shen, Phys. Lett. {\bf B236} (1989) 173-178; Nucl. Phys. {\bf 339B} (1990) 191-221; Phys. Lett. {\bf B242} (1990) 401-406; Phys. Lett. {\bf B245} (1990) 72-78

\bibitem{FKN2} M. Fukuma, H. Kawai, R. Nakayama, 
Comm. Math. Phys. {\bf 143} (1992) 371-403

\bibitem{BK} I. Bakas, E. Kiritsis, Int. J. Mod. Phys. {\bf A7} Suppl. {\bf 1A} (1992) 55-81

\bibitem{BKK} I. Bakas, B. Khesin, E. Kiritsis, Comm. Math. Phys. {\bf 151} (1993) 233-243

\bibitem{KR1} V.G. Kac, A. Radul, Comm. Math. Phys. {\bf 157} (1993) 429-457, hep-th/9308153

\bibitem{FKRN} E. Frenkel, V. Kac, A. Radul, W. Wang, Comm. Math. Phys. {\bf 170} (1995) 337-358, hep-th/9405121

\bibitem{Awata} H.~Awata, M.~Fukuma, Y.~Matsuo, S.~Odake,
Prog. Theor. Phys. Suppl. \textbf{118} (1995) 343-374,
hep-th/9408158

\bibitem{KR2} V.G. Kac, A. Radul, 
Transformation groups {\bf 1} (1996) 41-70, hep-th/9512150


\bibitem{SV}
  O. Schiffmann, E. Vasserot,
  Publications math´ematiques de l’IHES´ , {\bf 118} (2013) 213–342,
  arXiv:1202.2756

\bibitem{AS}
  N. Arbesfeld, O. Schiffmann,
  Symmetries, Integrable Systems and Representations (Iohara, Kenji and MorierGenoud, Sophie and R´emy, Bertrand, ed.), vol. {\bf 40} of Springer Proceedings in Mathematics and Statistics, pp. 1–13, Springer London, 2013

\bibitem{Tsim} A.~Tsymbaliuk,
Adv. Math. \textbf{304} (2017) 583-645,
arXiv:1404.5240

\bibitem{Prochazka} T.~Proch\'azka,
JHEP \textbf{10} (2016) 077,
arXiv:1512.07178

\bibitem{DI} J. Ding, K. Iohara, 
Lett. Math. Phys. {\bf 41} (1997) 181-193, q-alg/9608002

\bibitem{Miki} K. Miki, J. Math. Phys. {\bf 48} (2007) 123520

\bibitem{K} M. Kapranov,  
Algebraic geometry {\bf 7}, J.Math. Sci. {\bf 84} (1997) 1311-1360, alg-geom/9604018

\bibitem{BS} I. Burban, O. Schiffmann, 
Duke Math. J. {\bf 161} (2012) 1171, arXiv:math/0505148

\bibitem{S}  O. Schiffmann, 
J. Algebraic Combin. {\bf 35} (2012) 237-26, arXiv:1004.2575

\bibitem{Feigin} B. Feigin, M. Jimbo, T. Miwa, E. Mukhin,
Commun. Math. Phys. \textbf{356}  (2017) 285, arXiv:1603.02765

\bibitem{GJ} D. Goulden, D.M. Jackson, A. Vainshtein,
Ann. of Comb. {\bf 4} (2000) 27-46,
Brikh\"auser, math/9902125

\bibitem{MMN} A.~Mironov, A.~Morozov, S.~Natanzon,
Theor. Math. Phys. \textbf{166} (2011) 1-22,
arXiv:0904.4227\\
A.~Mironov, A.~Morozov, S.~Natanzon,
J. Geom. Phys. \textbf{62} (2012) 148-155,
arXiv:1012.0433

\bibitem{MMCal} A.~Mironov, A.~Morozov,
Phys. Lett. \textbf{B842} (2023) 137964,
arXiv:2303.05273

\bibitem{Jimbo} B. Feigin, M. Jimbo, T. Miwa, E. Mukhin, Kyoto J. Math. {\bf 52} (2012) 621-659, arXiv:1110.5310

\bibitem{MMZ} A.~Mironov, A.~Morozov, Y.~Zenkevich,
Eur. Phys. J. \textbf{C81} (2021) 461,
arXiv:2103.02508

\bibitem{GMTsPieri} D.~Galakhov, A.~Morozov, N.~Tselousov,
JHEP \textbf{08} (2023) 049,
arXiv:2307.03150

\bibitem{AT} B.~Azheev, N.~Tselousov,
Nucl. Phys. B \textbf{1018} (2025) 116975,
arXiv:2503.07583

\bibitem{KN} Anatol N. Kirillov, M. Noumi, Duke Math. J. {\bf 93(1)} (1998) 1-39, q-alg/9605004


\bibitem{Vinet} L.~Lapointe, L.~Vinet,
Commun. Math. Phys. \textbf{178} (1996) 425-452,
q-alg/9509003\\
L.~Lapointe, L.~Vinet, {\sl A Rodrigues formula for the Jack polynomials and the Macdonald-Stanley conjecture}, q-alg/9509002

\bibitem{Dunkl} Charles F. Dunkl, 
Trans. AMS, {\bf 311 (1)} (1989) 167-183

\bibitem{Ch} I. Cherednik,
  Vol. {\bf 319}. Cambridge University Press, 2005

\bibitem{DF} Vl. Dotsenko, V. Fateev, Nucl. Phys. {\bf B240} (1984) 312-348

\bibitem{Wak} M. Wakimoto, Comm.Math.Phys. {\bf 104} (1986) 605

\bibitem{FF} B. Feigin and E. Frenkel, UMN {\bf 43} (1988) 227

\bibitem{GMMOS} A. Gerasimov, A. Marshakov, A. Morozov, M. Olshanetsky and S. Shatashvili, Int.J.Mod.Phys. {\bf A5} (1990) 2495

\bibitem{CS} S.-S. Chern, J. Simons,
Ann.Math. {\bf 99} (1974) 48-69

\bibitem{Gua} E. Guadagnini, M. Martellini, M. Mintchev, Clausthal 1989, Procs.307-317; Phys.Lett. {\bf B235} (1990)
	275

\bibitem{MS} A.~Morozov, A.~Smirnov,
Nucl. Phys. \textbf{B835} (2010) 284-313,
arXiv:1001.2003

\bibitem{China1} R.~Wang, C.~H.~Zhang, F.~H.~Zhang, W.~Z.~Zhao,
Nucl. Phys. \textbf{B985} (2022) 115989,
arXiv:2203.14578

\bibitem{China2}
R. Wang, F. Liu, C.H. Zhang, W.Z. Zhao,
Eur. Phys. J. {\bf C82} (2022) 902, arXiv: 2206.13038

\bibitem{MMsc} A.~Mironov and A.~Morozov,
JHEP \textbf{03} (2023) 116,
arXiv:2210.09993

\bibitem{Kerov} S.V. Kerov, Func.An.and Apps. {\bf 25} (1991) 78-81

\bibitem{MMkerov} A. Mironov, A. Morozov, Journal of Geometry and Physics, {\bf 150} (2020) 103608, arXiv:1811.01184

\bibitem{Macrs} I.G. Macdonald, {\sl Orthogonal polynomials associated with root systems}, preprint (1987);
mathQA/0011046

\bibitem{Chrs} I. Cherednik, 
q-alg/9412016\\
I. Cherednik, 
The Annals of Mathematics, Second Series, {\bf 141} (1995) 191-216

\bibitem{Vogel} P. Vogel, {\sl Algebraic structures on modules of diagrams}, Preprint (1995), available at \url{https://webusers.imj-prg.fr/~pierre.vogel/diagrams.pdf}, J. Pure Appl. Algebra, {\bf 215} (2011) 1292-1339\\
P. Vogel, {\sl The Universal Lie algebra}, Preprint (1999), available at \url{https://webusers.imj-prg.fr/~pierre.vogel/grenoble-99b.pdf}

\bibitem{BM} L.~Bishler, A.~Mironov,
Phys. Lett. \textbf{B867} (2025) 139596,
arXiv:2504.13831

\bibitem{BMM} L.~Bishler, A.~Mironov and A.~Morozov,
Phys. Lett. \textbf{B868} (2025) 139695,
arXiv:2505.16569

\bibitem{MSl} A.~Morozov and A.~Sleptsov,
{\sl Vogel's universality and the classification problem for Jacobi identities},
arXiv:2506.15280

\bibitem{B} L.~Bishler,
{\sl Vogel's universality and Macdonald dimensions},
arXiv:2507.11414

\bibitem{paper:NS-cherednik}
  M.Nazarov,~E.Sklyanin,
  IMRN, {\bf 2019(8)} (2019) 2266–2294, arXiv:1703.02794



\end{thebibliography}
\end{document}